\documentclass[runningheads]{llncs}

\usepackage[T1]{fontenc}
\usepackage[utf8]{inputenc}
\usepackage[english]{babel}
\usepackage{amsmath,amssymb}

\usepackage{microtype}
\usepackage{bussproofs}
\usepackage{dsfont}
\usepackage{listings}
\usepackage{hhline}
\usepackage{caption}
\usepackage{subcaption}
\usepackage{mathtools}
\usepackage{booktabs}
\usepackage{tabularx}
\usepackage{multirow}
\usepackage{rotating}
\usepackage{makecell}
\usepackage{cite}
\usepackage{xcolor}
\usepackage{enumitem}
\usepackage{pgf}
\usepackage{float}
\usepackage{listing}

\usepackage{dafny}
\usepackage{listings-rust}
\usepackage{listings-verus}

\lstdefinestyle{DefaultStyle}
{
  basicstyle=\scriptsize\ttfamily\linespread{0.8}\selectfont,
  breaklines=true,
  tabsize=1,
  breakindent=2em,
  literate={\ \ }{{\ }}1
}

\lstdefinestyle{Verus}%
{ style=DefaultStyle,
, identifierstyle=%
, commentstyle=\color[gray]{0.4}%
, stringstyle=\color[rgb]{0, 0, 0.5}%
, keywordstyle={[1]\bfseries}%
, keywordstyle={[2]\color[rgb]{0.75, 0, 0}}%
, keywordstyle={[3]\color[rgb]{0, 0.5, 0}}%
, keywordstyle={[4]\color[rgb]{0, 0.5, 0}}%
, keywordstyle={[5]\color[rgb]{0, 0, 0.75}}%
, keywordstyle={[6]\color[rgb]{0, 0, 0.75}}%
, columns=spaceflexible%
, keepspaces=true%
, showspaces=false%
, showtabs=false%
, showstringspaces=true%
}%

\lstdefinestyle{VerusLineNos}{
  style=Verus%
, numbers=left%
, firstnumber=auto%
, numberblanklines=true%
, numberstyle=\color{gray}%
, numbersep=5pt%
, xleftmargin=15pt%
}

\newcommand{\tr}[1]{}                 %
\newcommand{\eat}[1]{}                %

\newcommand{\ProtocolFour}[4]%
{\begin{eqnarray*}\mathsf{Protocol:}&&#1\\&&#2\\&&#3\\&&#4\end{eqnarray*}}

\def\centerhack#1{\hbox to 0pt{\hss\footnotesize #1\hss}}
\def\dchack#1{\vbox to 0pt{\vss{\hbox to 0pt{\hss#1\hss}}\vss}}

\def\FullBox{\hbox{\vrule width 8pt height 8pt depth 0pt}}
\def\qed{\ifmmode\qquad\FullBox\else{\unskip\nobreak\hfil
\penalty50\hskip1em\null\nobreak\hfil\FullBox
\parfillskip=0pt\finalhyphendemerits=0\endgraf}\fi}

\makeatletter
\def\sectionautorefnamesymbol{\S\@gobble}
\def\sectionautorefname{\sectionautorefnamesymbol}
\def\subsectionautorefname{\sectionautorefnamesymbol}
\def\subsubsectionautorefname{\sectionautorefnamesymbol}

\makeatother

\newcommand{\goalref}[1]{\hyperref[#1]{\textbf{G\ref*{#1}}}}
\newcommand{\chalref}[1]{\hyperref[#1]{\textbf{C\ref*{#1}}}}

\newcommand{\taulocal}{\tau_{\text{local}}}

\newcommand{\eqdef}{\triangleq}

\newcommand{\capfig}[2]{\textbf{#1}. {\small \textit{#2}}}

\lstdefinestyle{DefaultStyle}
{
  basicstyle=\scriptsize\ttfamily\linespread{0.8}\selectfont,
  breaklines=true,
  tabsize=1,
  breakindent=2em,
  literate={\ \ }{{\ }}1
}

\lstdefinestyle{Verus}%
{ style=DefaultStyle
, basicstyle=\ttfamily\normalsize
, identifierstyle=%
, commentstyle=\color[gray]{0.4}%
, stringstyle=\color[rgb]{0, 0, 0.5}%
, keywordstyle={[1]\bfseries}%
, keywordstyle={[2]\color[rgb]{0.75, 0, 0}}%
, keywordstyle={[3]\color[rgb]{0, 0.5, 0}}%
, keywordstyle={[4]\color[rgb]{0, 0.5, 0}}%
, keywordstyle={[5]\color[rgb]{0, 0, 0.75}}%
, keywordstyle={[6]\color[rgb]{0, 0, 0.75}}%
, columns=spaceflexible%
, keepspaces=true%
, showspaces=false%
, showtabs=false%
, showstringspaces=true%
}%

\lstdefinestyle{VerusLineNos}{
  style=Verus%
, basicstyle=\ttfamily\scriptsize
, numbers=left%
, firstnumber=auto%
, numberblanklines=true%
, numberstyle=\color{gray}%
, numbersep=5pt%
, xleftmargin=15pt%
}

\lstdefinestyle{VerusLineNosCrStrike}{
  style=VerusLineNos%
, basicstyle=\scriptsize\ttfamily\linespread{0.8}\selectfont\color{red}%
, commentstyle=\color{red}%
, keywordstyle={[1]\color{red}}%
, keywordstyle={[2]\color{red}}%
, keywordstyle={[3]\color{red}}%
, keywordstyle={[4]\color{red}}%
, keywordstyle={[5]\color{red}}%
, keywordstyle={[6]\color{red}}%
, moredelim=[s][{\itshape\color{red}}]{\#[}{]}%
}

\newcolumntype{L}[1]{>{\raggedright\let\newline\\\arraybackslash\hspace{0pt}}m{#1}}
\newcolumntype{C}[1]{>{\centering\let\newline\\\arraybackslash\hspace{0pt}}m{#1}}
\newcolumntype{X}[1]{>{\raggedleft\let\newline\\\arraybackslash\hspace{0pt}}m{#1}}

\newcolumntype{R}[2]{%
    >{\adjustbox{angle=#1,lap=\width-(#2)}\bgroup}%
    l%
    <{\egroup}%
}

\usepackage{hyperref}
\usepackage{cleveref}

\begin{document}

\renewcommand{\sectionautorefname}{Section}
\renewcommand{\subsectionautorefname}{Section}

\let\subsectionautorefname\sectionautorefname
\let\subsubsectionautorefname\sectionautorefname

\title{Tunable Automation \\ in Automated Program Verification}

\titlerunning{Tunable Automation in Automated Program Verification}
\author{Alexander Y. Bai\inst{1,3}\orcidID{0009-0009-7458-7864}
\and
Chris Hawblitzel \inst{2}\orcidID{0000-0002-5676-0362} \and
Andrea Lattuada \inst{3}\orcidID{0000-0002-9303-452X}
}

\authorrunning{A. Y. Bai, et al.}

\institute{
New York University
\email{ayb5065@nyu.edu}
\and
Microsoft Research
\email{chris.hawblitzel@microsoft.com}
\and
MPI-SWS
\email{abai@mpi-sws.org, andrea@lattuada.me}
}

\maketitle

\begin{abstract}

Automated verification tools based on SMT solvers have made significant progress in verifying complex software systems. However, these tools face a fundamental tension between automation and performance when dealing with quantifier instantiation—the primary source of incompleteness and verification slowdown in SMT-based verifiers. Tools choose between aggressive quantifier instantiation that provides more automation but longer verification times, or conservative instantiation that responds quickly but may require more manual proof hints.

We present a mechanism that enables fine-grained control over the availability of quantified facts in verification contexts, allowing developers to selectively tune the level of automation. Our approach lets library authors provide different pre-defined automation levels while giving end-users the ability to further customize quantifier availability at the module, function, or proof context level.

We implement our techniques in Verus, a Rust-based verification tool, and evaluate them on multiple openly available codebases. Our empirical analysis demonstrates the automation-performance tradeoff and that selective quantifier management enables developers to select the appropriate level of automation in different contexts.


\end{abstract}

\section{Introduction}\label{sec:intro}

Formal verification of programs can guarantee their correctness with respect to a specification (and under certain assumptions of their operating environment). Verification tools based on automated theorem provers (such as SMT-solvers) have made significant progress in efficiency and scalability to larger projects~\cite{veribetrkv,verus-sosp24,verismo,anvil,verified-storage}.
Fully automated proofs are only possible when writing specification and code so that it falls within a decidable fragment of first-order logic \cite{liquid-haskell, leon, predictable-verification, serval, ivy-pldi18}, but this imposes a significant limitation in expressivity and often prevents the developer from describing the specification and program in the most natural way~\cite{verus-sosp24}. Program verifiers thus allow expressing programs and proofs in an undecidable logic, often first-order logic with uninterpreted functions, integer arithmetic, and unrestricted quantification.

\begin{figure}[h]
    \centering
    \includegraphics[width=0.9\linewidth]{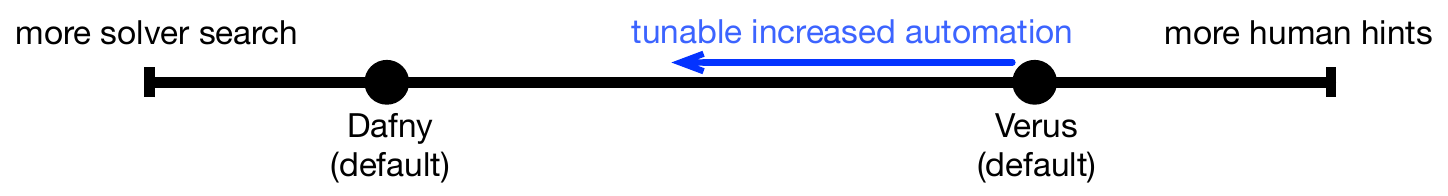}
    \caption{The spectrum of quantifier-based automation.
    \label{fig:spectrum}}
\end{figure}

The verification community's experience is that quantifier instantiation is the primary source of incompleteness and long verification time in SMT-based verifiers~\cite{veribetrkv, viper, trigger16, verus-oopsla23, prog-with-triggers}. These tools express everything from core axioms to developer-stated properties using quantifiers, and then rely on the SMT-solver syntactic matching (e-matching) heuristic \cite{ematching}, guided by \emph{triggers}, to automatically instantiate quantified expressions with symbolic values from the proof context. This approach has two consequences: (i) if the syntactic trigger that results in instantiation is not present in the proof context, the user has to provide a hint to instantiate the quantifier as needed and (ii) the verification time can be proportional, often exponentially \cite{prog-with-triggers}, to the number of quantifiers in the automated prover's scope and to how easily they are instantiated.
If quantifiers are not sufficiently instantiated, the verification tool's automation is more incomplete and requires additional manual hints for verification success. On the other hand, quantifiers with easily matched syntactic triggers may enable the prover to automatically find proofs in more cases, but result in a large potential search space, causing long verification runtimes before the tool reports success or failure. 
Semi-automated verification tools generally fall somewhere on this ``automation spectrum'' (\autoref{fig:spectrum}). For example, Dafny~\cite{leino2010dafny} employs a trigger selection algorithm that minimizes the need for user annotation and hints: this helps new Dafny users write proofs without immediately having to learn about the underlying mechanism~\cite{trigger16}. On the contrary, Verus~\cite{verus-oopsla23,verus-sosp24} has conservative defaults optimized for short response times, at the potential cost of less out-of-the-box automation due to more incomplete instantiation.

In this work, we explore several ways to allow the user to tune the amount of automation (and its resulting cost) provided by quantifier instantiation in semi-automated verification tools. We implement and evaluate our techniques in the Verus Rust-based verification tool and language\cite{verus-oopsla23, verus-sosp24}.

Our technical contribution is a \verb|broadcast| mechanism to selectively include axioms and lemmas in the proof environment as quantified facts, allowing both coarse-grained and fine-grained control over which quantifiers are available to the SMT solver in different proof contexts. Quantified facts can be imported in bulk, or selectively, at the module, function, or isolated proof context level. The \verb|broadcast| mechanism enables library authors to provide various default ``automation levels'' and end-users to select among these, or further tune the proof context. The user can also query the tool to determine which quantified facts were likely used in a proof, allowing fine-tuning once a proof is discovered.

We provide an empirical analysis of the \verb|broadcast| mechanism's impact on verification time and proof burden across multiple openly available Verus codebases. Our evaluation clarifies the automation tradeoff that allows shorter proofs at the cost of some verification performance. Additionally, we investigate trigger selection strategies and their effects on the automation-performance tradeoff.

\section{Background}
\subsection{Verus}

Verus \cite{verus-oopsla23, verus-sosp24} is an SMT-based semi-automated tool to verify the correctness of code written in Rust. Verus is heavily inspired by Dafny~\cite{leino2010dafny} and optimizes for short developer iteration time due to short response times for both successful and unsuccessful proofs~\cite{verus-sosp24}. The following Verus example briefly demonstrates Verus syntax and how it integrates a verification language with Rust:

\begin{lstlisting}[language=Verus,style=VerusLineNos]
spec fn divides(n: int, k: nat) -> bool { n %

spec fn is_prime(n: nat) -> bool {
  forall|k: nat| 2 <= k < n ==> !divides(n as int, k)
}

spec fn is_even(i: int) -> bool { divides(i, 2) }

proof fn even_gt_2_isnt_prime(i: nat)
  requires i > 2 && is_even(i as int)
  ensures !is_prime(i) { }

fn is_prime_impl(n: u64) -> (result: bool)
  requires n >= 2,
  ensures result == is_prime(n as nat)
{ /* ... implementation and proof ... */ }
\end{lstlisting}

The \verb|spec| functions \verb|divides|, \verb|is_prime|, and \verb|is_even| contain pure logical ghost expressions (including quantifiers), used to express function specifications or auxiliary conditions. \verb|even_gt_2_isnt_prime| is a ghost \verb|proof| function, which expresses a callable lemma, and - in this case - can be proven automatically by the solver. \verb|is_prime_impl| is an executable function that is proven to compute whether the input is prime according to the \verb|is_prime| definition (we omit the implementation and proof for brevity).
Ghost code includes \verb|spec| and \verb|proof| functions, executable functions' pre- and post-conditions, and inline assertions, such as the following, which appears as part of the proof of \verb|is_prime_impl|:
\begin{lstlisting}[language=Verus,style=VerusLineNos]
assert(divides(n as int, i as nat));
\end{lstlisting}

\subsection{Quantifier Triggers} \label{sec:triggers}

As discussed in \autoref{sec:intro}, semi-automated verification tools heavily rely on quantifier instantiation to automatically discharge proof obligations. Syntactic matching (e-matching) via ``triggers''~\cite{simplify} is a commonly used heuristic for automatic quantifier instantiation, and -- while the underlying solver is often capable of automatically selecting quantifier triggers (sometimes known as ``patterns") -- languages like Dafny and Verus perform trigger selection at the surface language level and allow manual selection and manual overrides through user-provided annotations. Performing trigger selection in the verification tool %
can help avoid expensive or vacuous triggers \cite{trigger16}.

In Verus, like in similar tools, only function calls, field accesses, or arithmetic operators where one of the arguments is one of the quantified variables are allowed as trigger subexpressions. A trigger is composed of potentially multiple subexpressions and needs to mention all quantified variables at least once.

The following proof function has a precondition that all elements of a sequence are even numbers, and we want to prove that a specific entry is even.

\begin{lstlisting}[language=Verus,style=VerusLineNos]
spec fn is_even(i: int) -> bool { i %

proof fn seq_trigger_example(s: Seq<int>)
    requires
        5 <= s.len(),
        forall|i: int| 0 <= i < s.len() ==> #[trigger] is_even(s.index(i)),
{
    assert(s.index(3) %
}
\end{lstlisting}

The valid triggers for the \verb|forall| on line 6 are \lstinline{is_even(s.index(i))} and \lstinline{s.index(i)}, and the user has manually chosen \lstinline{is_even(s.index(i))} with the \verb|#[trigger]| annotation. Because \lstinline{is_even(s.index(3))} does not appear anywhere in the context, the solver will not instantiate the quantifier with \verb|i == 3| and thus the \verb`assert` on line 8 will fail. Changing the trigger to \verb|s.index(i)| will enable matching with \verb|s.index(3)| and result in a successful proof.

\subsection{Implicit Context}\label{sec:implicit-context}

SMT-based semi-automated tools like Dafny and Verus include a default set of quantified facts provided as part of the ``prelude'' of the SMT problem that encodes the program verification condition. This default set defines, among others, the semantics of built-in concepts like sequences (\verb|Seq| in Dafny). For example, proving the following

\begin{lstlisting}[language=Verus,style=VerusLineNos]
proof fn seq_axiom_usage(s1: Seq<nat>, s2: Seq<nat>)
  requires s1.len() > 10 && s2.len() > 20
  ensures s1.add(s2).len() > 30 { }
\end{lstlisting}

relies on this built-in Verus axiom:

\begin{lstlisting}[language=Verus,style=VerusLineNos]
\end{lstlisting}

In Dafny, this context is fixed and defined as part of verification condition generation (through Boogie~\cite{boogie}). In Verus, a default context is defined in a pure-SMT prelude and a collection of lemmas in the standard library explicitly marked to become contextual quantified facts.
More quantified facts in the proof context improve the level of automation available to the users, but increase the proof search space available to the solver, potentially resulting in longer response times.
In this work, we address the one-size-fits-all approach to contextual quantified facts by allowing users to construct and selectively import them.

\section{Quantified Facts \`a la Carte}

We introduce a mechanism to selectively publish quantified facts from \verb|proof| functions (lemmas) with a user-provided marker (\verb|broadcast|) and import them as additional context for a module or function (via the new keyword \verb|broadcast use|). We implemented the \verb|broadcast| mechanism in Verus, but the general mechanism we introduced of manipulating quantified facts is independent of Verus, and can be applied to other program verifiers. 

We showcase the usage of \verb|broadcast| in the following code examples.

This lemma fails to verify automatically in Verus:

\begin{lstlisting}[language=Verus,style=VerusLineNos]
pub proof fn push_contains(a: Seq<int>) {
  let b = a.push(3);
  assert(b.contains(3));
}
\end{lstlisting}

due to Verus' default context being insufficient to prove the fact that a sequence contains every element it previously contains after pushing:

\begin{lstlisting}[language=Verus,style=VerusLineNos]
pub proof fn lemma_seq_contains_after_push<A>(s: Seq<A>, v: A, x: A)
    ensures (s.push(v).contains(x)) <==> v == x || s.contains(x),
{ /* ... manual proof ... */ }
\end{lstlisting}

However, once proven and marked with \verb|broadcast|, this lemma can be imported as the following universally quantified statement:

\begin{lstlisting}[language=Verus,style=VerusLineNos]
        (#[trigger] s.push(v).contains(x)) <==> v == x || s.contains(x)
\end{lstlisting}

As broadcast proofs share the same ability as moving the input parameters to a universally quantified variable in the ensures clause, they must adhere to trigger rules as if they are universally quantified. Therefore, the broadcasted version of \verb|lemma_seq_contains_after_push| will be the following, with the trigger on the syntactic expression \verb|s.push().contains()|:
\begin{lstlisting}[language=Verus,style=VerusLineNos]
pub broadcast proof fn lemma_seq_contains_after_push<A>(s: Seq<A>, v: A, x: A)
    ensures (#[trigger] s.push(v).contains(x)) <==> v == x || s.contains(x),
\end{lstlisting}

Now, this broadcasted proof can be imported in context with the directive: 

\begin{lstlisting}[language=Verus,style=VerusLineNos]
broadcast use {lemma_seq_contains_after_push};
\end{lstlisting}

For example, the initial lemma verifies automatically by importing:

\begin{lstlisting}[language=Verus,style=VerusLineNos]
pub proof fn push_contains(a: Seq<int>) {
  broadcast use {lemma_seq_contains_after_push};
  let b = a.push(3);
  assert(b.contains(3));
}
\end{lstlisting}

\emph{Broadcastable quantified facts}, i.e. \verb|proof| functions marked with \verb|broadcast|, can be defined in the Verus standard library, in third-party libraries, or in end-user code, and can be imported directly, or through \verb|broadcast groups|, which name and collect related broadcastable facts together so they can be imported in context in bulk.
As an example, \verb|lemma_seq_contains_after_push| is now in fact a broadcastable quantified fact in the Verus standard library, and it is part of the \verb|group_seq_properties| broadcast group defined, again, by the standard library:

\begin{lstlisting}[language=Verus,style=VerusLineNos]
pub broadcast group group_seq_properties {
    lemma_seq_contains,
    // ...
    lemma_seq_contains_after_push,
    // ...
}
\end{lstlisting}

This enables the user to import a larger set of quantified facts in context in bulk, as follows:

\begin{lstlisting}[language=Verus,style=VerusLineNos]
pub proof fn push_contains(a: Seq<int>) {
  broadcast use {vstd::seq_lib::group_seq_properties}; // imports all the quantified facts from the group
  let b = a.push(3);
  assert(b.contains(3));
}
\end{lstlisting}

Thanks to this mechanism Verus can provide a default set of quantified facts in context while allowing the end-user to tune the level of quantifier-instantiation-induced automation in each proof context. Library authors can also leverage the mechanism to provide (sets of) contextual quantified facts for their users' benefit.

\subsubsection{Restricting the Set of Quantified Facts in the Context}\label{sec:axiom-usage-info}

Additional quantified facts in context improve automation at the cost of some verification performance, as we will see in our case studies in \autoref{sec:exploring}. We extended Verus to interpret the solver's UNSAT-CORE output, which contains the facts that were used in a successful proof, to display to the user the set of imported quantified facts that are likely necessary for the proof. When running Verus with the appropriate flag on the last example of the previous section (which relies on the bulk import of the quantified facts in \verb|group_seq_properties|), we get the following output:

\begin{lstlisting}[basicstyle=\scriptsize\ttfamily]
checking this function used these broadcasted lemmas and broadcast groups:
        - (group) vstd::seq_lib::group_seq_properties,
        - vstd::seq_lib::lemma_seq_contains_after_push
\end{lstlisting}

This indicates that \verb|lemma_seq_contains_after_push| was the only quantified fact relevant for the proof, and enables the end-user to reduce the context by replacing the \verb|broadcast use| for the entire \verb|group_seq_properties| with just \verb|lemma_seq_contains_after_push|, which results in a successful proof but with reduced context:

\begin{lstlisting}[language=Verus,style=VerusLineNos]
pub proof fn push_contains(a: Seq<int>) {
  broadcast use {vstd::seq_lib::lemma_seq_contains_after_push};
  let b = a.push(3);
  assert(b.contains(3));
}
\end{lstlisting}

The combination of bulk import of quantified facts via \verb|broadcast use| of \verb|broadcast group|s and the tooling to report those that are likely relevant to the proof enables a workflow where the developer first imports a broad set of facts in context, and when the proof is successful, trims the set of quantified facts to those relevant to the proof, potentially recovering proof performance by reducing the solver's search space.

The ability to import quantified facts in broader or more limited proof scopes also enables fine-grained control of where to increase automation. With \verb|broadcast use|, a Verus user can import quantified facts in any proof contexts, or for an entire module. New proof contexts are introduced by
\begin{itemize}[noitemsep,topsep=2pt,parsep=0pt,partopsep=0pt]
\item proof functions (\verb|proof fn|);
\item \verb|assert (expression) by { /* ... proof ... */ }|;
\item calculational proofs (``calc'' in Dafny and Verus) \cite{dafny-calc}. %
\end{itemize}

In the previous example, the user can \verb|broadcast use group_seq_properties| at the module level, and then fine-tune the level of automation by tailoring the set of imported quantified facts in more granular proof contexts to address verification performance slowdowns~\cite{axolocl}.

\section{Broadcastable Quantified Facts in Practice}

\subsection{Modularizing Lemmas Associated to User-defined Abstractions}\label{ironkv}

The \verb|broadcast| mechanisms enable constructing self-contained abstractions that hide the internal details of type-specific proofs but expose relevant facts relating values in the abstraction. As an example, we look at the abstraction over keys used in IronKV, the Verus port~\cite{verus-sosp24} of the verified key-value store from IronFleet~\cite{ironfleet}.

For most of the system, the concrete type representing a \textit{key} is irrelevant for the proofs, which only rely on the fact that keys have certain properties. IronKV defines a Rust trait (similar to a Haskell typeclass) that (i) expresses the proof obligations for a type to be a valid \textit{key}, and (ii) exposes the properties relevant to the clients of the \textit{key} abstraction through the \verb|broadcast| mechanism.

\begin{figure}
\begin{lstlisting}[language=Verus,style=VerusLineNos]
pub trait Key {
    ...
    proof fn key_obligations()
        ensures // ... conditions necessary for the type to be a valid key

    broadcast proof fn trans_lt_lt(a:Self, b:Self, c:Self)
        ensures a < b && b < c ==> a < c
    { /* justified thanks to the ensures of `key_obligations` */ }

    // ... additional properties ...
}

pub broadcast group group_key_cmp_properties {
    Key::trans_lt_lt,
    // ... additional `broadcast` proofs from the trait
}
\end{lstlisting}
\caption{
  \label{fig:broadcast-cmp-properties}
  \capfig{Excerpt from the definition of IronKV's \textit{key} trait: the example has been simplified for ease of presentation but without hiding any relevant details}{}
}
\end{figure}

\autoref{fig:broadcast-cmp-properties} shows the definition of the \verb|Key| trait. A type implementing the trait must provide a proof for \verb|key_obligations|, which ensures that the type has the necessary properties to be used as a \verb|key|. In return, the trait exposes a number of broadcastable quantified facts (such as \verb|trans_lt_lt|) which are grouped together in the \verb|broadcast group| \verb|group_key_cmp_properties|. Code that operates on keys need not know the details of the type used as keys, and can just rely on the properties exposed through the group by importing the quantified facts in context via the directive:
\begin{lstlisting}[language=Verus,style=VerusLineNos]
broadcast use {group_key_cmp_properties};
\end{lstlisting}

\subsection{Modularizing Verus Standard Library}

As discussed in \autoref{sec:implicit-context}, built-in abstract datatypes' semantics are defined as axioms or proofs in the default implicit context of quantified facts. The \verb|broadcast| mechanism offers greater flexibility in organizing the quantified facts for these datatypes, such as sequences, maps, and sets.

Dafny fully defines these datatypes' semantics as axioms in the Boogie prelude used during verification condition generation: some can be derived from other axioms with proofs, but they still appear as axioms in the prelude because Dafny lacks a general mechanism to include in the context additional quantified facts derived from proofs. In comparison, in Verus, there are a much smaller number of axioms (resulting in a smaller Trusted Computing Base). The additional facts relating operations on the built-in datatypes are \verb|broadcast| \verb|proof| functions, and are grouped together with the axioms in a default \verb|broadcast group|. This \verb|broadcast group| is then automatically included in the proof context with the Verus' standard library.
As a conservative choice to prioritize verification performance, this is a smaller set than what is included in Dafny's prelude.

\section{Exploring the Automation Tradeoff}\label{sec:exploring}

We set out to study the effect of tuning the level of automation associated with reasoning about built-in Verus collection datatypes: \verb|Seq|, \verb|Map|, \verb|Set|, and \verb|Multiset|. We design experiments to answer the following questions:

\begin{enumerate}[label=\textbf{RQ\arabic*:}, labelwidth=3em, leftmargin=*, align=left, nosep]
    \item Does increasing the number of quantified facts in context result in more automation, i.e. fewer manual user-provided hints (in the form of \verb|assert|s)?
    \item Does increasing the number of quantified facts in context hinder verification performance or the verification experience? %
\end{enumerate}

We grouped the \verb|broadcast proof| functions corresponding to the quantified facts included in Dafny's prelude but excluded from Verus' default context in four \verb|broadcast group|s, one per collection type: \lstinline{group_<type>_properties}. The following table lists the number of translated Dafny prelude lemmas:

\smallskip

\begin{center}
\begin{tabular}{|l||c|c|c|c||c|}
\hline
\textbf{Collection datatype} & \textbf{Seq} & \textbf{Map} & \textbf{Set} & \textbf{MultiSet} & \textbf{Total} \\
\hline
\# of \texttt{group\_<type>\_properties} lemmas & 25 & 2 & 10 & 12 & 49 \\
\hline
\end{tabular}
\end{center}

We then used \verb|broadcast use| to include these lemmas as quantified facts in the context of a number of openly available Verus projects and measured the impact on the number of manual hints necessary and on verification performance.
In the following, we refer to imported quantified facts for collection types as \emph{ambient facts} for brevity.
For these experiments, we focus on the effects of including these facts in all proof contexts, and do not try to fine-tune the scope in which facts are made available. All experiments are available at: 
\href{https://github.com/ahuoguo/tunable-automation}{https://github.com/ahuoguo/tunable-automation}

\subsubsection{Projects under Study}\label{projects}
We considered the following as verification benchmarks:
\begin{itemize}[noitemsep,topsep=2pt,parsep=0pt,partopsep=0pt]
    \item IronKV is a distributed key-value store which was described in Sec~\ref{ironkv}
    \item Splinter\cite{splinter} is an ongoing work on a key-value store designed around a $B^\epsilon$ tree \cite{betree}.
    \item Anvil\cite{anvil} is a framework for building and formally verifying Kubernetes controllers. It provides a TLA-style temporal logic to reason with eventually stable reconciliation of controllers. We only focus on the Anvil framework rather than the verification of specific controllers.
    \item CapybaraKV\cite{verified-storage} is a storage system targeting persistent memory devices
\end{itemize}

\subsection{Minimization}
For the purpose of this experiment, we need a measure of ``automation'', for which we use a proxy metric: the number of hints provided to the solver via \verb|assert| statements that are needed for the proof to succeed.
Our projects under study contain more \verb|assert|s than strictly necessary: they are used during proof development as a proof debugging strategy, and developers do not systematically remove \verb|assert|s that become unnecessary once the proof is complete. \verb|assert|s are also often intentionally retained as documentation of the proof structure.
For the number of \verb|assert|s to be a viable proxy metric for the level of automation, we need to determine the least number of such hints that still result in successful verification. We can automate such minimization by removing \verb|assert|s and re-running verification, electing to keep only those that -- when removed -- result in a failure. However, this has combinatorial complexity: for a project with $ n $ \verb|assert|s, there are, as a first approximation, $ 2^n $ candidates for a ``minimal'' proof.
\footnote{Consider a verification project with 100 asserts you want to minimize, and each run is 1 second. Then you have $2^{100}$ possible candidates to consider, which takes about $4 \times 10^{22}$ years.}
We developed a Verus minimization tool that linearly scans through the proof code and removes the proof code (in the form of \verb|assert| and \verb|assert (...) by {...}|) if its removal does not cause a failure (i.e. if it is redundant as a hint to the solver). We consider this an acceptable approximation to make it feasible to use this metric.
Despite this simplification, it is Verus' performance that enables the use of this metric, for which we still need to re-run verification after each \verb|assert| is tentatively removed\footnote{Additional tooling may enable only re-running verification for the verification condition affected by the \lstinline{assert}, but there is ``cross-talk" between VCs (also known as ``instability'') which complicates this potential approach}.
To our knowledge, this is the first study of what fraction of \verb|assert|s in a project are actually necessary hints to the solver. In a sense, we introduce a form of program slicing in the context of verification code\cite{slicing}.

To evaluate the impact of ambient facts on the level of automation, we run our Verus minimization tool twice on each project: first on the unmodified project, and then again after importing the ambient facts. The number of \verb|assert|s that become redundant when the ambient facts have been imported is a proxy metric for the resulting increase in automation (and thus, indirectly, proof effort).

It is important to note that counting the number of surviving \verb|assert|s after ``minimization'' is a very rough measure of automation and might not reflect true developer experience.
Developers use \verb|assert|s to guide the solver, but also to debug proofs (and to document and clarify their own understanding of the proof strategy). Another limitation is our ``minimizer'' only targets assertions, namely, \verb|assert(...)|,  \verb|assert <forall> ... by {...}|, while it ignores calls to lemmas (\verb|proof| functions) which can also be made redundant by ambient facts\footnote{As Verus allows proof function to return a value, deleting lemma calls requires more sophisticated mechanisms.}. %

\subsection{Effect on the Level of Automation}

Table~\ref{tab:min-results} contains the effect of ambient facts. One interesting observation is the number of extra assertions the target verification projects have: more than half of the assertions in these projects do not contribute to the SMT solver achieving a successful verification.

\begin{table}[h]
\centering
\begin{tabular}{|l|c|c|c|}
\hline
\textbf{System} & \textbf{Original \#} & \textbf{Minimized} & \textbf{Minimized with} \\
& \textbf{of asserts} & \textbf{(baseline)} &\textbf{Ambient Facts} \\
\hline
IronKV   & 646       & 268   &     245(-23, -8.6\%)            \\
Splinter &   2678    & 1158     &     1130(-28, -2.4\%)           \\
Anvil    &   701    &  343   &     336(-7, -2.0\%)            \\
CapybaraKV & 941       & 449    &     415 (-34, -7.6\%) \\
\hline
\end{tabular}
\caption{Assertion counts for verification systems after minimization and importing ambient facts}
\label{tab:min-results}
\end{table}
\vspace{-2mm}
\noindent \textit{RQ1: Do more ambient facts about collection types bring more automation?}

We see a reduction of the number of assertions for ambient facts, ranging from 2\%-9\%. Some of the assertions are non-trivial proof blocks ranging from 5-10 lines and were automatically discharged solely due to the ambient facts of collection types.

\subsection{Impact on Verification Time}\label{sec:impact-of-broadcasting}

\begin{figure}[h!]
    \centering
    \resizebox{1\linewidth}{!}{\input{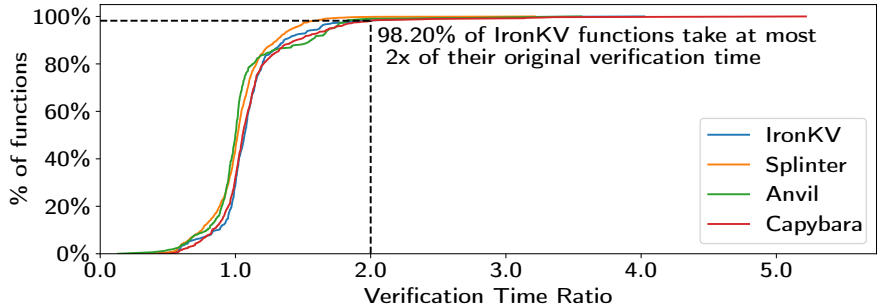}}
    \caption{Cumulative distribution of verification time ratio between ambient facts (minimized) and original verification time for each function. We removed two extreme cases of 10x+ verification slowdown, one in Splinter, where the runtime bumped from 34ms to 669ms (19.1x), and one in Anvil, where the runtime bumped from 3508ms to 43563ms (12.4x).}
    \label{fig:all}
\end{figure}

To answer RQ2, we first measure the ambient facts' impact on verification time. All experiments are done with 9 threads on a Dell PowerEdge M620 blade system (Intel Xeon E5 v2) with 16 cores using Verus version 0.2025.07.03.3105aa2. %
The results can be seen in \autoref{fig:all}. The figure compares each function's original verification time to the verification time after having ambient facts and minimization. We can see that ambient facts do slow down verification, with $\sim2\%$ of the functions taking 2x of their verification time, and the worst slowdown in all projects ranges from 3x to 19x.

The experiment data show that ambient facts improve automation, but may have noticeable runtime slowdown in one or two specific functions. We argue that our new technique for broadcastable quantified facts is a way to combat verification slowdowns on particular functions. Quantified facts can be tuned down by importing them at a finer-grained level, rather than providing the ambient facts everywhere, as we discussed in \autoref{sec:axiom-usage-info}.

\vspace{-2mm}

\subsection{Impact on Verification Failure Time}\label{sec:sample-failure}

\begin{figure}[h!]
    \centering
    \resizebox{1\linewidth}{!}{\input{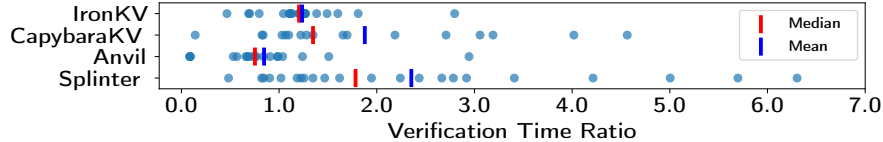}}

    \caption{Runtime Ratio for 20 runs (each with one of the randomly sampled asserts removed) against successful verification.}
    \label{fig:failure-sample}
\end{figure}

Although Verus's run time for successful verification tasks is important when we are importing ambient facts, the time for reporting a verification failure is also important to the verification engineer's experience. As we may have dramatically increased the search space by importing more ambient facts, this might cause a similarly dramatic increase in the time Verus takes to report a verification failure. Once the solver finds a proof, it stops, which does not happen if the proof is incomplete or incorrect. Thus, it is also important to see if more ambient facts impact user experience by slowing down the reporting of failures. After minimization, removing an assertion will almost certainly result in verification failure. Therefore, we randomly sampled 20 assertions to remove and recorded the time it took to report the failure, and computed the ratio to the successful verification time for the function left intact. The results can be seen in \autoref{fig:failure-sample}. We conclude that ambient facts did result in slowdowns in a small fraction of functions that failed to verify, with 78\% of the failed functions still verifying in 2x of their original verification time. Thus, ambient facts did not result in an explosion in verification failure time.

\vspace{-2mm}

\section{Automation from Triggers}

Quantified facts are instantiated via syntactic matching of ``triggers'' (\autoref{sec:triggers}). When the verification tool selects the ``trigger'' expressions for a quantifier, it can default to selecting every possible valid, non-redundant trigger, or it can conservatively only select \emph{safe} triggers, i.e. those that are less likely to significantly increase the search space that the solver can explore, leading to longer verification times.

\subsection{Trigger Selection Strategy}

Verus' default automatic trigger selection is quite conservative: it selects a few ``safe'' triggers. A quantifier can be annotated with \verb|#![all_triggers]|, which switches the trigger selection strategy to pick all valid, non-redundant triggers for the quantified expression\footnote{Due to technical reasons related to polymorphism, there is an exception: quantifiers where the quantified variables are used as arguments to both arithmetic operations and function calls. For these quantifiers, the 
\lstinline{#![all\_triggers]} strategy remains more conservative}. The \verb|#![all_triggers]| strategy is generally still more conservative than Dafny's default trigger selection strategy (which employs some more advanced techniques where quantifiers are automatically split and triggers shared, see Sec 2.1 of \cite{trigger16}).

\subsection{More Automation with More Triggers}

\subsubsection{Axioms}

An immediate question follows: Is it possible to obtain a more powerful automation (that is, considering our metric, a smaller number of required hints to the solver as \verb|assert|s) by changing how triggers are selected for the default set of quantified facts imported when using the Verus standard library?

The answer is no. By manual inspection of most axioms of Verus standard library's collection datatypes, \lstinline{Seq, Set, Map, Multiset}, we have observed that their triggers are quite straightforward, and most unselected trigger candidates would generally lead to ``matching loops" which result in a very large search space and solver timeouts \cite{trigger16}. For example, for the following axiom of lengths of sequences (an excerpt of the default set of the Verus standard library's ambient facts):

\begin{lstlisting}[language=Verus,style=VerusLineNos]
pub broadcast axiom fn axiom_seq_add_len<A>(s1: Seq<A>, s2: Seq<A>)
    ensures
        #[trigger] s1.add(s2).len() == s1.len() + s2.len(),
;
\end{lstlisting}

The two possible trigger sets are 

\begin{lstlisting}[language=Verus,style=VerusLineNos]
\end{lstlisting}

But the second one is almost always unhelpful, as this axiom is usually only relevant in contexts where there is a call to  \lstinline{add}. Moreover, if the second trigger is included, this axiom will be instantiated excessively: once on each pair of \lstinline{.len()} calls in the program and proof context.

\subsubsection{User Code}
We saw that there are few opportunities to expand the triggers of quantified facts in the standard library, but user-level code may contain more complex quantified expressions with more trigger candidates. In this section, we study the effects of selecting a broader set of triggers for all quantifiers.

In the following experiment, we picked IronKV and removed all manual triggers if possible. We then used \lstinline{#![all_triggers]} wherever it is supported, and ran the \verb|assert| minimization tool. The change only breaks one proof. In the end, we added 206 \verb|#![all_triggers]| annotations in the 261 quantifiers in IronKV.

\begin{figure}[htbp]
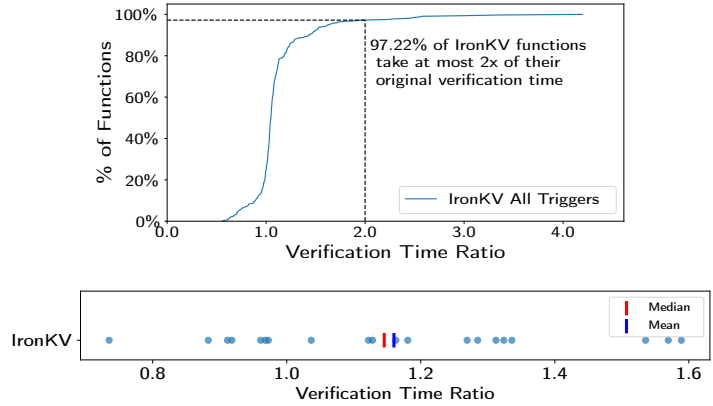

    \centering

    \begin{subfigure}[b]{0.6\textwidth}
        \centering
        \resizebox{1\linewidth}{!}{\input{figures/ironkv_at.pgf}}
    \end{subfigure}

    \begin{subfigure}[b]{0.8\textwidth}
        \centering
        \resizebox{1\linewidth}{!}{\input{figures/ironkv_at-sample.pgf}}
    \end{subfigure}
    \caption{Cumulative distribution of verification time, and verification ``slowdown" ratio for 20 sampled asserts for IronKV with \lstinline{all_triggers} enabled. Detailed descriptions of the two figures can be seen in \autoref{fig:all} and \autoref{fig:failure-sample}.}
    \label{fig:ironkv-at}
\end{figure}

We measure the impact on automation using our \verb|assert|-based metric, and report results in \autoref{tab:min-results-at}.
For this project, a more liberal trigger selection strategy improves automation (according to our metric) more than importing ambient facts for collection types. We measure the performance impact (on success and failure), with a methodology similar to the one described in \autoref{sec:impact-of-broadcasting} and \autoref{sec:sample-failure}, and report the results in \autoref{fig:ironkv-at}.

\begin{table}[h!]
\centering
\begin{tabular}{|l|c|c|c|c|c|}
\hline
\textbf{System} & \textbf{Original } & \textbf{Min } & \textbf{All Triggers} & \textbf{Ambient Facts} \\
\hline
IronKV   & 646       & 268    &     235(-33, -12.3\%) & 245(-23, -8.6\%) \\
\hline
\end{tabular}
\caption{Assertion counts for IronKV after minimization with all\_triggers}
\label{tab:min-results-at}
\end{table}
\vspace{-4mm}

Unfortunately, this experiment cannot scale well to other verification projects considered in \Cref{projects}, as they have at least twice as many triggers throughout the project, and verifying the project using \verb|all_triggers| pervasively resulted in a large amount of verification failures. We suspect these failures are due to increased search space and misguided heuristics. However, our data indicated that more trigger candidates on the user-level quantified expressions have a non-trivial positive impact on proof automation. We leave further exploration of the automation level of different trigger levels as future work.

\vspace{-2mm}

\section{Related Work}

\subsubsection{Automation Provided in Other Program Verifiers}

The authors of Dafny argue that trigger-based instantiation should be available at the source code instead of relying on SMT heuristics, and provide a trigger selection algorithm\cite{trigger16}.

Increasing automation is crucial for past large-scale verification projects in Dafny. One way of publishing additional quantified facts pervasively on a module level is to declare a \lstinline{ghost predicate} in a module with an \verb|ensures| clause. The following is an example of publishing the fact that the uninterpreted function \lstinline{BinOp} is commutative.

\begin{lstlisting}[language=Dafny]
  function BinOp(x: int, y: int) : int

  ghost predicate BinOpAuto(x:int, y: int)
    ensures BinOpAuto(x, y) ==> BinOp(x, y) == BinOp(y, x)
    // BinOpAuto is equivalent to  
    //  forall x,y. BinOpAuto(x, y) ==> BinOp(x, y) == BinOp(y, x)
\end{lstlisting}

By importing the module containing this \verb|ghost predicate|, the quantified fact derived from the \verb|ensures| is effectively available on the module level.

The authors of the VeribetrKV crash-safe key-value store~\cite{veribetrkv, linear-dafny} report that this methodology has been crucial in their development, and similar automation techniques appear in Why3 \cite{why3} (eg: lemma functions). However, this approach lacks granular control over where the quantified facts are imported (in contrast to \verb|broadcast proof|).

Axolocl is a tool to automatically localize proof context in local proof blocks \cite {axolocl}. Their proof localization algorithm moves relevant axioms for verifying an assertion into a local proof context, based on information from UNSAT-CORE and quantifier instantiation, in Boogie\cite{boogie}, an intermediate verification language for Dafny and Viper. We believe that, with Verus proof blocks and broadcastable quantified facts, we can achieve source-level translation to minimize proof contexts.

\subsubsection{Instability}

Instability (or brittleness) is a common phenomenon observed in SMT-based verification tools. Namely, some proofs break under semantically equivalent changes\cite{mariposa, dafny-brittle}. Shake \cite{shake} showed that irrelevant context is one of the culprits for instability, and proposed an SMT-level instability mitigation strategy. Multiple SMT-based tools are starting to address instability \cite{cazamariposa, conjecture, cvc5-instability}.

A technique known as \emph{free facts} was proposed in Dafny to fight instability. free facts is a set of generated facts based on syntactic matching during the verification condition generation\cite{freefacts}. In a sense, free facts perform a weaker version of pattern-based instantiation before SMT solving. The free facts authors indicate a strong interest in studying their verification projects with a reduced number of assertions, as the extra assertions take up verification time and resource usage. Our empirical evaluation on Verus suggests that reducing the number of assertions does not necessarily reduce verification time, and we leave exploration in this direction as future work. 

\subsubsection{Proof Automation in Verus}

RagVerus \cite{ragverus} is the first work that considered repository-level proof automation. Though the methodologies are very different from this project, their proof automation aspect is very similar to our evaluation of quantified facts in \autoref{sec:exploring}.

ProofPlumber\cite{proofplumber} introduced several automated strategies to insert assertions in the right places to guide the proof engineer to see where a proof has gone wrong. It is an interesting future work to incorporate the minimizer and axiom-usage-info as automatic strategies.

\section{Conclusion}

We explored several directions for tunable automation in the Verus Rust verification tool. To this end, we introduced \emph{broadcastable quantified facts} as a mechanism to tune automation on a finer-grained level, and demonstrated its usefulness through case studies. We also conducted multiple experiments on openly available large Verus projects to evaluate the cost and benefits of increasing automation along two axes: increasing quantified facts in context, and increasing the level of automation from triggers. 

\section*{Acknowledgements}
\label{sec:Acknowledgements}

We thank Jialin Li, Xudong Sun, and Zhizhen Cai for their help with the case studies. We also thank Alexandre Moine, Bryan Parno, and the audience of the Verus Retreat 2025, FP Teatime, NYU FM seminar, and VSTTE 2025 for their helpful feedback. Pranav Srinivasan initially wrote \verb|all_triggers| in Verus. The first author especially thanks Jay Bosamiya, Travis Hance, and Bryan Parno at CMU REUSE 2023 for getting him involved in the Verus world.

\bibliographystyle{splncs04}
\bibliography{references}

\appendix
\newpage
\section{Appendix}

\subsection{Mariposa}

One interesting experiment is to use Mariposa\cite{mariposa} to answer whether more ambient facts introduce more ``instability", partly answering RQ2. This is not included in the main paper as Verus programs tend to be more ``stable'' than other semi-automated verification tools. Nevertheless, \autoref{tab:mariposa} contains the results of running Mariposa on the four projects. Except for Anvil having a reverse trend, most projects have a slight increase in instability when exposed to a larger set of ambient factors, which aligns with our expectations.

\begin{table}[ht]
\centering
\small
\begin{tabular}{l cc cc cc c}
\toprule
\multirow{2}{*}{\textbf{Benchmark}} & \multicolumn{2}{c}{\textbf{Stable}} & \multicolumn{2}{c}{\textbf{Unstable}} & \multicolumn{2}{c}{\textbf{Unsolvable}} & \multirow{2}{*}{\textbf{Total}}\\
 & \textit{Number} & \textit{\%} & \textit{Number} & \textit{\%} & \textit{Number} & \textit{\%} & \\
\midrule
IronKV Original        & 363      & 100.00\% & 0   & 0.00\% & 0   & 0.00\% & 363  \\
IronKV Minimized       & 363      & 100.00\% & 0   & 0.00\% & 0   & 0.00\% & 363  \\
IronKV Broadcasted     & 362      & 100.00\%  & 0   & 0.00\% & 0   & 0.00\% & 363  \\
IronKV AllTriggers     & 362      & 99.72\%  & 1   & 0.28\% & 0   & 0.00\% & 363  \\
IronKV AllTriggers Minimized  & 362 & 99.72\% & 1  & 0.28\% & 0   & 0.00\% & 363  \\
Splinter Original      & 1458     & 99.05\%  & 5   & 0.34\% & 9   & 0.61\% & 1472 \\
Splinter Minimized     & 1441     & 97.89\%  & 21  & 1.43\% & 10  & 0.68\% & 1472 \\
Splinter Broadcasted   & 1442     & 97.96\%  & 20  & 1.36\% & 10  & 0.68\% & 1472 \\
Anvil Original         & 315      & 99.68\%  & 1   & 0.32\% & 0   & 0.00\% & 316  \\
Anvil Minimized        & 311      & 98.42\%  & 5   & 1.58\% & 0   & 0.00\% & 316  \\
Anvil Broadcasted      & 314      & 99.37\%  & 2   & 0.63\% & 0   & 0.00\% & 316  \\
CapybaraKV Original    & 723      & 99.86\%  & 1   & 0.14\% & 0   & 0.00\% & 724  \\
CapybaraKV Minimized   & 719      & 99.31\%  & 5   & 0.69\% & 0   & 0.00\% & 724  \\
CapybaraKV Broadcasted & 721      & 99.59\%  & 3   & 0.41\% & 0   & 0.00\% & 724  \\
\bottomrule
\end{tabular}
\caption{Mariposa Results on All Projects, ``Broadcasted" means minimized with ambient facts, described in \autoref{tab:min-results}}
\label{tab:mariposa}
\end{table}

\subsection{Implementation}

The underlying implementation uses \emph{fuels} to bring the broadcasted proof into proof context \cite{dafny-fuels}. If a proof function is marked as \verb|broadcast|, then it will create a broadcast version that moves the parameters to be universally quantified. If \verb|broadcast use| appears, Verus will activate the fuel such that the broadcasted proof function can be used, thus bringing the broadcasted proof function in scope.

Timeline: \href{https://github.com/verus-lang/verus/commit/bc1fe84bcf0dd596c0f806d90a4ad08a1c41f842}{Verus@bc1fe} provided an initial implementation; however, only proof functions without a proof body (mostly axioms) can be treated as ambient facts, as at the time we cannot check if there are no cycles in the lemmas. The following PR \href{https://github.com/verus-lang/verus/pull/850/}{Verus\#850} implements SCC order for SMT commands to make SMT prove the broadcasted proof functions before they get imported globally. Finally, \href{https://github.com/verus-lang/verus/pull/1022}{Verus\#1022} implements the generalized broadcastable quantified facts.

\end{document}